# Evaluating Design Conformance Through Trace Comparison

Reid Anderson and Hassan Reza
University of North Dakota
{reid.b.anderson, hassan.reza}@und.edu

***Abstract*—The design of a system and its implementation are two tasks often carried out by different individuals on a development team, and can occur weeks or months apart. This creates a potential for divergence between real behavior and the designed model an implementation is intended to match. Particularly as time passes and individuals that were present for the original conception of the design leave, a system can lose coherence and drift from intended design principles. Even with a robust design for a system, more is needed to guarantee that the key details of the implementation match the design, and that adherence to a particular strategy is not lost over time. This paper proposes an approach to address that concern for distributed systems using conformance checking, a methodology borrowed from process mining. Distributed traces produced by instrumented applications are evaluated for conformance by comparison to design traces. The resulting conformance percentage is a quantitative metric that can be tracked over time to determine how closely a concrete implementation corresponds to the key attributes of the expected design model. This analysis is done using the dominant industry standard, OpenTelemetry, and so should be applicable to a wide range of distributed systems.**

## I. Introduction

Distributed systems present a variety of benefits across a wide range of industries. Each node can be independently deployed or updated and may use substantially different hardware or software. This allows for rapid growth, quick iteration, and flexibility as the needs of the system change and older technologies become obsolete. [1] The development process for a distributed system allows multiple teams to explore various directions simultaneously, but this flexibility presents significant challenges for maintaining system oversight and architecture in a complex environment. This paper proposes a process to address a single research question:

**Research Question:** How can an individual industry practitioner quantitatively verify that a distributed system design is implemented by the running application?

A robust answer to this research question would allow for confident decision-making based on the design model. To represent the core functionality expected from a process evaluating the comparison between design and reality for a distributed system and allowing an individual to carry it out, we define 5 requirements for the system below:

1) Require minimal change to running applications.
2) Allow an individual practitioner to apply the methodology without organizational change, and without risk to the organization.
3) Perform at scale when applied widely to complex systems across an organization.
4) Provide visually clear feedback of success or failure.
5) Provide a quantitative metric that can be tracked and improved over time.

We propose a framework using distributed tracing, telemetry data commonly collected from distributed systems. The traces observed in the running system will be compared to defined design traces that describe required and disallowed system behavior. We show how these design traces can be defined, compared with observed behavior, and finally how the results of the comparison can be evaluated and used to drive system improvement.

Our approach borrows heavily from *conformance checking*, a sub-discipline of process mining most often applied to manufacturing or business processes. [2] We draw an analogy between a distributed software system and a complex manufacturing process taking place in an industrial plant. There was a detailed plan for the construction of the plant, the result of which is the ability to produce output through the manufacturing process. However, after initial construction has finished and production has started, portions of the process will change over time. The requirements for the finished product will change, particular inputs may become unavailable, or efficiency benefits from reconfiguration may be discovered. Regardless of the reason, it is critical that there is a model which accurately reflects the actions being taken in the factory. This allows engineers to plan, identify issues, and evaluate the performance of the industrial plant.

The similarities with a complex software system are strong. Such a system begins from a design, the implementation of which provides the ability to service user needs. These user needs and the requirements of the system are certain to change over time after initial construction. This is particularly true as faster, iterative development has been the constant direction of travel in software engineering for many years at this point. [3] It is still critical that a model exists which accurately reflects the actions being taken by the system. Without an accurate model, software engineers cannot make informed decisions.

## II. Background

Conformance checking is driven by two primary inputs: a process model and event logs. Event logs are produced by a process as it executes, indicate the actions performed by the system, and come in a variety of forms and formats. Their important characteristic is that they accurately reflect in substantial detail the actions taken by the system. For an iteration of the process that *conforms*, two things are true. First, all required operations in the model are performed in reality and represented in event logs. Second, no operations that are explicitly disallowed by the model are performed or present in event logs.

In an industrial setting, the event logs that drive the comparison to the process model are produced by machines or software supporting the process, and have a variety of data formats. Examples of conformance checking in literature include medical treatment and manufacturing plants. [4] [2] Beyond case studies, much of the conformance checking literature is concerned with statistical measures for the results. [5]

We adopt distributed traces as the event logs that will drive our conformance checking process. A trace gives a description of all actions taken by multiple services or components as a request is processed. [6] Each individual action taken in a trace is known as a *span* and is reported by the application performing the operation a span represents. A *trace* is simply a collection of these spans. The benefits of distributed tracing were first realized at scale in the late 2000s and early 2010s by large technology companies, with Dapper being one of the first high volume deployments described in the literature. [6] Dapper was used by Google to observe and monitor its search infrastructure and was able to perform at massive volumes. This idea caught on and began to be adopted by many large technology companies operating distributed systems. [7] Eventually a standardization movement emerged and in 2019 competing standards were merged into OpenTelemetry as the industry standard for distributed tracing and telemetry collection. [8] The use of an industry standard as well as existing widespread collection of trace data means that the necessary event log information to perform conformance checking already exists in many organizations.

## III. Related Work

There has been some work to apply conformance checking concepts to software engineering, but the focus is primarily where formal or high detail specifications reliably exist - which are relatively uncommon in distributed architectures. [9] Specifically, *conformance testing* has been applied to software systems to guarantee the correctness of an implementation. [10] Our proposed process attempts to accomplish similar aims. It should be viewed as building upon existing conformance testing work to apply to distributed systems without a formal specification.

The primary use case that organizations gather traces for is to monitor systems. But, there has long been a recognized benefit to the insight that traces provide about the system. The most well-developed field in trace analysis is of system performance over time. This has been a rich area for machine learning and big data approaches towards spotting trends and detecting anomalies in running applications. [7] The most common use case is to identify issues as they develop, prior to user reports. Once the anomaly has been detected, the supporting trace information should make it straightforward for engineers to identify the problem. In addition to real-time support, there has been work to identify long-running sources of performance degradation in the system. [11] Areas that most significantly bottleneck the throughput of the system are identified, which helps guide prioritization of work to improve the performance of the system.

Trace analysis has also been done outside of system performance. Two primary areas are in privacy and security, where audits of trace data can be considered superior to source code audits. [6] The complexity of distributed systems means that data is often passed in a chain ending far from the database where it originated. This is concerning when working with sensitive information like financial data or health records. Controls on the source database may be strong, but information that diffuses throughout the system may become exposed in areas with fewer controls. Frameworks have been developed that use traces to follow the chain of sensitive data, and can evaluate and reduce the level of data risk. [12] Data maintainers can more effectively control access and reduce the possibility of data leaks. Similarly, analysis of traces and telemetry data allows for identification of security risks. This can be based on deviations from known traffic patterns or accessing of sensitive resources by unexpected parties. [13] The privacy and security analysis of traces touches closest to the topics covered in this paper, because they compare the system as it is to an idealized state. However, a privacy and security analysis of the system is purely concerned with ensuring actions are taken safely. The evaluation of similarity between plan and reality, which is not substantially developed in the distributed tracing community at the moment, is the focus of the remainder of this paper.

## IV. Methodology

Our approach is centered around the definition of custom design traces, and the comparison of design traces to observed traces. A *design trace* is a trace, intended to convey the expected execution behavior of the system, where each span has been defined instead of collected from an executing application. These spans are created by the designer of the system, ideally before implementation has begun of the enhancement or system under consideration. An *observed trace* on the other hand, is a trace that was emitted by the running system and has been collected for future analysis.

Our implementation of conformance checking for design verification is a static process consisting of the below steps performed on a collected data set, using design traces as the process model, and observed traces as the event logs.

1) Define design traces
2) Collect observed traces
3) Perform conformance checks
4) Analyze results and identify issues

### A. Define Design Traces

The first step in producing a design trace is to define the required or disallowed operations as spans and specify how they are connected. Traces can be represented as a directed acyclic graph (DAG), so to visualize the design trace as it is defined, we use a *span graph*. [11] A *span graph* is a graphically displayed DAG where spans are represented as nodes and parent/child or link relationships between spans are represented as edges. A *child* span represents a sub-operation of the parent span, and each child span may have at most one parent. Links represent causal relationships between two spans that may be part of the same or different traces, and a span may have zero-to-many links.

Observability visualization is a developed field, with many commercial software offerings. The use of a span graph to reflect the structure of execution paths taken by the application should be readily familiar to practitioners that have worked with observability data in the past, and help engineers in constructing design traces that accurately reflect a system.

New systems can be described using consistent documentation and design philosophies from the beginning. In this case, spans for each operation are manually defined.

TABLE I. Design Span Attributes

| Attribute | Description |
|---|---|
| design.description | Human-readable description. |
| design.maxDuration | Max duration in $\mu$s of a span. |
| design.allowNonImmediateParent | If true, check full parent tree. |
| design.isDisallowed | If true, matching is prohibited. |

A *root span*, a span with no parent, is the entry point to the operation. The root span is created, and then each subsequent child or linked span added until the desired execution path is defined. This continues until the new system represents what the designer expects and is ready for implementation. An observed trace can also be imported from a running system. After an initial import, spans can be modified or removed, to simplify the design trace to the critical path of operations that should be verified.

After the structure of the design trace has been completed, detailed properties for each span need to be specified. Each OpenTelemetry span has a defined collection of attributes, which are key-value pairs containing arbitrary information about the operation being described. Common use cases with predictable properties of interest, such as spans of HTTP requests, often have a consistent set of attributes defined. [14] We treat spans in design traces similarly, and Table I gives the consistent custom attributes we have defined. When defined, these attributes are used in the conformance checking process and allow for customization of evaluation behavior.

The completed design traces gives us the process model that conformance checking requires. All required operations (where *design.isDisallowed* is false) are defined, as well as all disallowed operations (where *design.isDisallowed* is true). From this point, the implementation of the system should proceed and, once completed, observed traces can be collected from the system for comparison.

### B. Collect Observed Traces

As described above, one of the strongest practical arguments for design verification based on distributed traces is that in many large scale distributed system deployments, traces are already collected and stored for other purposes. An export of observed trace data from the observability provider is required. A wide variety of commonly used platforms exist, but the general pattern is the same. The overall population of traces is filtered down to those that should be validated against the design traces across a particular time frame, converted to the OpenTelemetry format, and then exported to a static dataset.

### C. Perform Conformance Checks

The two inputs required to perform conformance checking are now present. We have defined the design traces

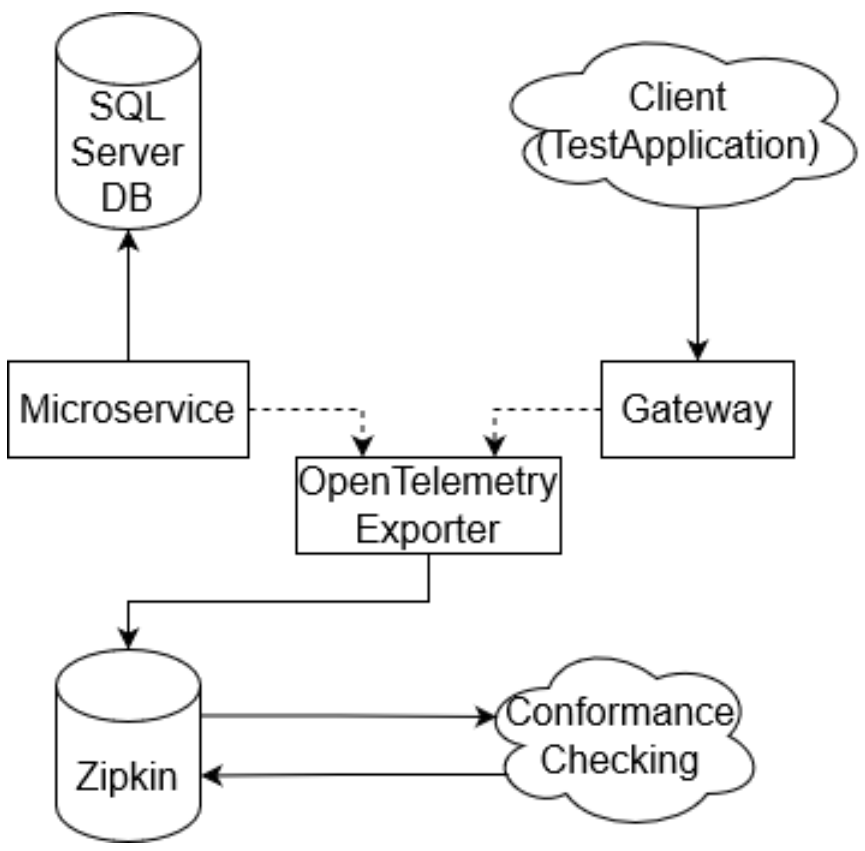

Fig. 1. A microservice system used to demonstrate the conformance checking process.

(likely a handful) that represent the process model and obtained the observed traces (likely many thousands) which are the event logs demonstrating the actions taken by the system. The task is then to iterate over each observed trace, and determine if it conforms to the criteria specified by the design trace. The evaluation for each observed trace is completely independent, and so processing the collection of observed traces is easily parallelized.

The conformance of a particular observed trace is determined as follows. Spans in design traces are split into two collections, the collection of required spans that must be present, and the collection of disallowed spans that must be absent. Each required span is evaluated to ensure a corresponding span is present in the observed trace and any conformance attributes (Table I) specified are satisfied. The evaluation that required attributes are present and that the criteria specified by the *design.** attributes are satisfied is trivial and straightforward, properties of the observed span are just compared to the expected values defined on the design trace. Slightly trickier is making sure that the parent span defined by the design trace is satisfied by the observed span. This behavior is what guarantees that the structure of the observed trace matches the structure of the design trace. First, the *design.allowNonImmediateParent* attribute is evaluated – if false or not provided the observed parent span is compared to the design parent span for conformance. The child span is conformant if they match, otherwise not. When design.allowNonImmediateParent is true, this comparison is done recursively until the root span is reached. If a span anywhere in that chain is found that matches the parent design span, the child is conformant – otherwise it is not. If a trace is marked non-conformant, we also store the span identifier from the relevant design span that was not satisfied and from the observed trace if applicable.

Each disallowed span is then evaluated to ensure a corresponding span is *not* present in the observed trace. The same logic is applied as for the required trace, but if an observed span is found which matches the design trace, the observed trace is considered non-conformant.

### D. Analyze Results and Identify Issues

As we described briefly in the background on conformance checking, there are a large variety of measures that can be used to evaluate the adherence of observed behavior to the process model. Our goal in this paper is to drive action for practicing software engineers, and so we focus on a basic and easily understandable evaluation metric. We define *conformance percentage* as a single, simple, output metric calculated as the number of conformant observed traces divided by the number of observed traces. This gives a straightforward measure of how closely the implementation adheres to the design, and the metric should be easily understandable to anyone working with the system. Tracking the conformance percentage over time should reveal any divergence between design and implemented functionality.

The second piece of output produced by the proposed process are the span identifiers that resulted in a non-conformant observed trace. This gives specific feedback as to the source of failures in design validation. Individual non-conformant traces can be examined, the specific failure highlighted, and overlaid with the expected operation from the design trace.

## V. Results

To illustrate the application of conformance checking to a distributed system and demonstrate the proposed methodology, we designed, implemented, and instrumented a sample microservice system as a case study. Three deviations from the specified design were intentionally introduced as part of the implementation, impacting a percentage of overall requests. A test application was written to feed network requests into the system, and adherence to the design was measured by conformance checking.

Figure 1 shows the sample microservice system. The test application, gateway API, and microservice were implemented using ASP.NET with a Microsoft SQL Server database. Each application was instrumented with OpenTelemetry .NET SDK libraries and collected telemetry data was stored in Zipkin. The core functionality of the system is a client calling a gateway API, which uses a single microservice to save and retrieve data from a relational database. This design is simple, but there are several design requirements to validate:

1) *The gateway API should never communicate directly with the database.* A core principle of microservices is that operations against data owned by a microservice should go through its provided contract. [15]

TABLE II. Span Attributes and Properties for System Design Traces

| traceId/spanId | Trace 1/Span A | Trace 1/Span B | Trace 1/Span C | Trace 2/Span D | Trace 2/Span E |
|---|---|---|---|---|---|
| name | aspnet_core.request | aspnet_core.request | sql_server.query | aspnet_core.request | sql_server.query |
| service.name | gateway | microservice | microservice | gateway | gateway |
| parentSpanId | null | Span A | Span B | null | Span D |
| design.description | Client request | Process request | DB operation | Client request | DB operation |
| design.maxDuration | 500 ms | null | null | null | null |
| design.allowNonImmediateParent | false | true | true | true | true |
| design.isDisallowed | false | false | false | true | true |

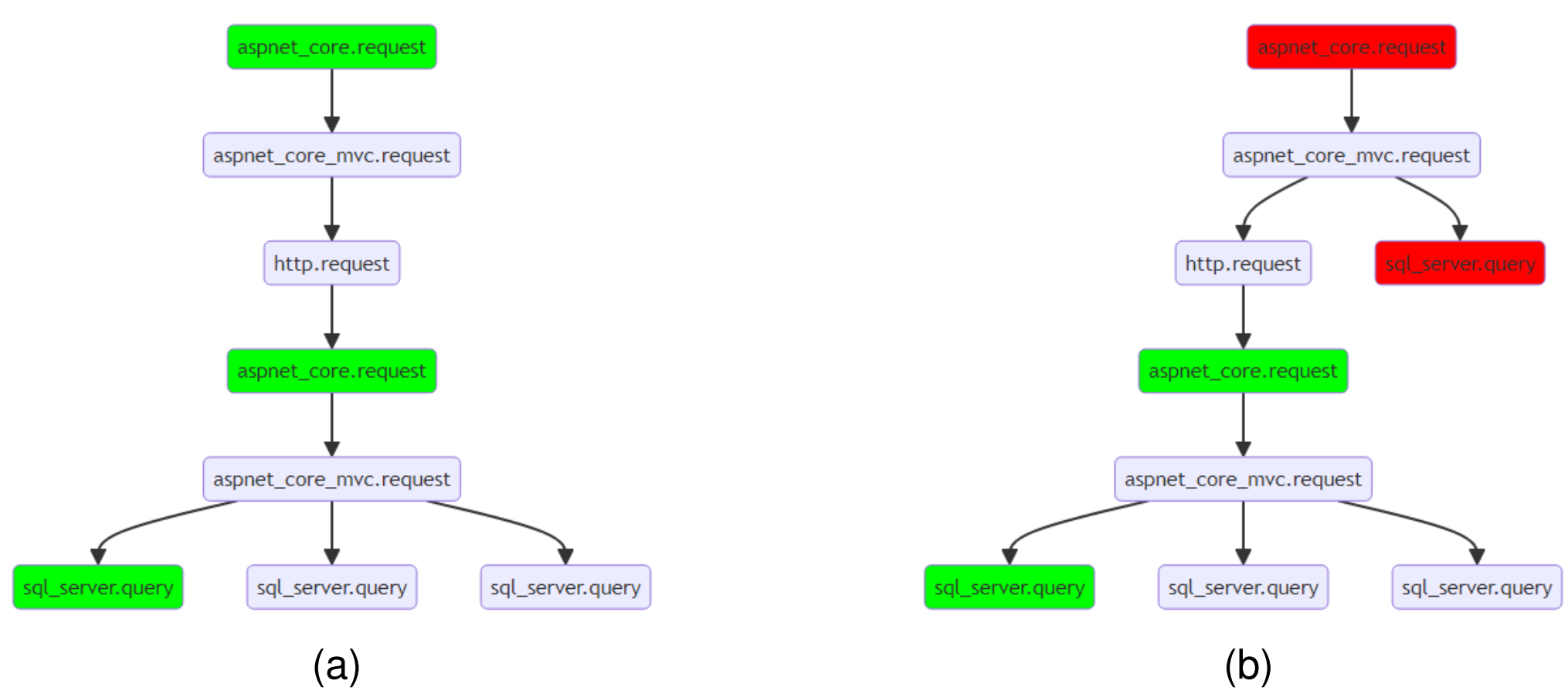

Fig. 2. Span graph for two observed traces. Subfigure a is conformant, subfigure b is non-conformant.

2) *Gateway* → *Microservice* → *Database*, should be the call order for every request.
3) *The overall duration of the operation should be less than 500 milliseconds.*

The first step in the process is the definition of the design traces. In order to satisfy the design requirements, there is one design trace with required spans, and one with disallowed spans. In addition to span relationships, the properties, and attributes are shown in Table II. The duration requirement is enforced by the *design.maxDuration* value given for Span A. The three required spans enforce that the correct call ordering is performed and the disallowed spans ensure that no database operation is performed by the gateway.

After implementation of the system, data was collected for analysis. The test application was used to send 100,000 requests to the gateway API. The observed traces from these requests were collected into Zipkin. Using the API endpoints provided by Zipkin, the traces were retrieved, converted to an OpenTelemetry compliant JSON format, and saved to the filesystem. With defined design traces and a dataset of observed traces, the inputs necessary to perform conformance checking are present. The observed traces were distributed to multiple threads and checked in parallel using the methodology described above. The conformance of each trace was determined, and if found to be non-conformant, the design spans that were violated recorded.

Deviations from the design were intentionally introduced to the implemented system, and the conformance checking process was able to correctly identify and categorize the errors. Out of 100,000 requests, 81,088 were found to be conformant and 18,912 non-conformant for one or more reasons. The conformance percentage for the system as a whole was 79.9%, with failures due to the omission of required spans (6,985 spans), durations longer than the specified maximum (5,982 spans), and the presence of disallowed operations (7,445 spans). This measurement allows improvement to be tracked over time, and provide concrete evidence for when a system has been brought fully into conformance with its design.

It is also useful to analyze the individual observed traces for insight into how a system is running compared to the expected design. Two observed traces from the implemented system are shown in Figure 2. The most apparent difference between the designed and observed traces is that there are considerably more observed spans. These were automatically collected by the OpenTelemetry instrumentation library to provide additional information, and is typical as traces are intended to give detailed insight. The conformance checking process is robust to additional information, as it checks minimum requirements and prohibited operations. The span graph in Figure 2a shows a conformant observed trace. All required spans are present, and no disallowed spans are observed — the trace is conformant.

There were multiple failures in observed behavior that resulted in non-conformant traces. Figure 2b shows an observed trace that failed to conform. The gateway service directly accesses the database, a violation of design principles and an operation that was defined as disallowed. Additionally, this request took longer than the 500 millisecond requirement to process. The required operation at the root of the trace is therefore not present, because it fails the necessary duration criteria. We now revisit how our original objective:

**Research Question:** How can an industry practitioner quantitatively verify that a distributed system design is implemented by the running application?

Traces are already commonly collected for observability purposes, so no change should be needed to running applications. Likewise, the definition of design traces do not require changes to existing artifacts or the installation of custom software. A practitioner is able to undertake this analysis without additional system or organizational change. They can evaluate the performance of the system as a whole using metrics like conformance percentage and clearly see sources of success or failure for specific traces in visualizations like Figure 2.

This satisfies four out of the five requirements laid out in the introduction to this paper. Further analysis of the process is needed to ensure that the methodology performs when scaled up to a system handling massive volumes of trace data. The 100,000 traces that were processed as part of the case study is a small sample compared to the scale that modern observability platforms operate at, so additional scalability analysis is needed. [6]

## VI. Conclusion and Future Work

Applying conformance checking to software gives insight into the similarity between design expectations and implementation. This paper describes a relatively simple approach, but one that should reveal opportunities to improve system maintainability and design conformance.

There are two primary areas for future research. Applying the process on a larger scale, and using more sophisticated statistical techniques to measure the results. Millions or billions of traces are produced by even moderately sized distributed architectures, and so the scalability of our analysis methods need to be proven at that scale. Then, the conformance checking research community has done a substantial amount of work on producing statistical measures to evaluate how well a process model reflects reality. These are significantly more detailed than the simple percentage measurement that was performed in this paper. Identifying measures that are still readily understood by practitioners and more precisely reflect the relationship between the observed and design traces would be valuable.

Modern software systems are often complex distributed environments with a large amount of concurrent change occurring. Methods that allow that flexibility while providing guardrails to ensure strong agreement between conceptual understanding and reality are valuable. The ability of conformance checking to quantitatively measure compliance in complex processes, and the demonstration that those concepts are transferable to a distributed system, is a strong argument for further research and application.